\documentclass[aps,prl,twocolumn,superscriptaddress,showkeys,10pt]{revtex4-1}
\usepackage{amsmath}
\usepackage{amssymb}
\usepackage{cancel}
\usepackage{graphicx}
\usepackage{esint}
\usepackage[latin1]{inputenc}
\begin{document}
\title{Realizing tunable non-Hermitian skin effects in dynamical quantum systems via the relative phase between multiple time-periodic driving}
\author{Huan-Yu Wang}\email{eurake27hywang@fzu.edu.cn}
\affiliation{Fujian Key Laboratory of Quantum Information and Quantum Optics and \\
Department of Physics, Fuzhou University, Fuzhou 350108, China}
\begin{abstract}
We demonstrate how the relative phase between the multiple time periodic driving can decide the emergence and the favorable localization direction of non-Hermitian skin modes. For the static non-Hermitian quantum chain with parity time symmetry, non-Hermitian skin effects (NHSEs) can be prohibited. As the dynamical driving is turned on, NHSEs get artificially reactivated, where the relative phase can serve as the controlling switch by breaking the temporal symmetry constraints. Meanwhile, a change of relative phase can also alter the spatial structures of the long-time averaged  effective Hamiltonian, which will consequently lead to the variation of skin localization direction for systems of higher dimensions. Our formalisms can be generally realized in diverse optical and mechanical platforms, and will pave the way for realizing tunable skin density profiles.
\end{abstract}
\maketitle
\textit{Introduction.---} One of the most intriguing topics in the realm of open quantum systems can be appointed as non-Hermitian skin effects (NHSE), where all the bulk eigen modes are accumulated at the boundary of the lattice \cite{PhysRevLett.121.086803,PhysRevLett.116.133903,Nat2,PhysRevLett.128.157601,Commun.Phys.5.252,PhysRevA.104.022215,PhysRevA.106.052216,PhysRevB.109.155137,xvv6-44yh,PhysRevB.99.245116}. Typically, the formation of NHSE is caused by the non-reciprocal tunneling or onsite dissipations \cite{PhysRevLett.77.570,PhysRevLett.133.076502,PhysRevLett.123.016805,vhz9-xwf4,bmq5-7tf6,PhysRevLett.125.186802}, and from the perspective of quantum transport, NHSE can be indicated by the non-zero area enclosed by the momentum space complex energy spectrum \cite{PhysRevLett.125.126402, PhysRevLett.124.086801}. Meanwhile, for the non-Hermitian system with topological nontrivial properties, the presence of NHSE may hinder the application of conventional bulk-boundary correspondence, where the proper topological characterizations should be constructed based on the redefined complex Bloch wave vector \cite{PhysRevLett.121.136802,PhysRevLett.121.026808,PhysRevX.8.031079,PhysRevLett.125.226402,PhysRevB.107.035101,PhysRevLett.125.226402,PhysRevB.107.195112,PhysRevLett.123.066404,PhysRevX.9.041015,Quantum.Sci.Technol.9.025019,PhysRevLett.123.066404,PhysRevResearch.1.023013,PhysRevLett.124.056802,YeXiong,PhysRevResearch.5.033058,Natphys}.

Beyond the results so far, time-periodic driving can be conceived to provide extra physical dimensions, which will lead to more fruitful features of dynamical NHSE \cite{PhysRevLett.133.070801,chk3-crdn,Nat.Phys.16.761,NatCommun2.2,PhysRevLett.128.120401,PhysRevB.108.035107,PhysRevLett.132.063804,PhysRevB.109.184302,NatCommun2.3,chk3-crdn,PhysRevResearch.6.023004,PhysRevA.109.063329,PhysRevResearch.6.023081,PhysRevLett.133.073803,s3b6-wz16,PhysRevB.107.165401,Science2,PhysRevLett.127.270602} and  exotic topological phase transitions \cite{PhysRevB.82.235114,PhysRevX.3.031005,Nat.Rev.Phys,qn87-bm33,PhysRevB.95.195155,PhysRevLett.125.183001,PhysRevB.96.195303,PhysRevB.105.214305,PhysRevB.90.205108,PhysRevB.101.235403,PhysRevB.101.235403,PhysRevLett.124.057001,PhysRevLett.127.067001}. Correspondingly, there will exist more diverse approaches to synthesize artificially controllable skin density patterns.

In this work, we demonstrate how NHSE can be switched off (or on), by tuning the relative phase between the  multiple time-periodic driving, and during which process, the presence of on-site dissipation or non-reciprocal quantum tunneling is always maintained. Specifically, with the non-zero windings of the momentum space energy spectrum defined as the point gap topological invariants to identify NHSE, the relative phase fosters the capability of giving birth to topological phase transitions by breaking the temporal symmetry constraints of the Floquet evolving operator. This procedure will equivalently destroy or reactive NHSE. Such a formalism can be applied when the dynamical Hamiltonian at each instantaneous time possesses parity time (PT) symmetry, and considering the cases of time-reversal symmetry or particle hole symmetry, a change of relative phase can no longer switch the NHSE. Meanwhile, for non-Hermitian systems of higher dimensions, there may exist favorable localization direction of the skin modes, which is decided by the spatial structure of the model or boundary shape of the lattice. A variation of relative phase can also alter the spatial structure of the long time averaged effective Hamiltonian, and thus changes the skin localization direction. Considering the dynamical driving in reality, the two formalisms above can be combined together to artificially design the skin density patterns. Finally, our results are feasibly to be simulated in diverse platforms exhibiting Floquet dynamics.

\begin{figure*}[t]
	\centering
	\includegraphics[width=0.97\textwidth,height=0.19\textheight]{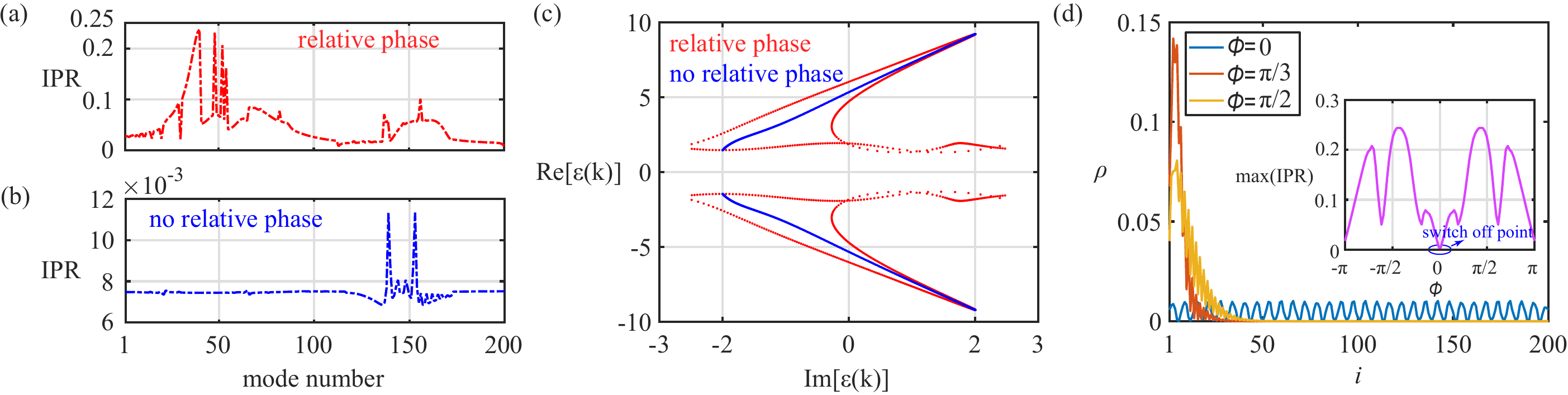}
    \caption{(a) For the dynamical system with multiple time-periodic driving, when the  relative phase $\phi=\frac{\pi}{2}$, and $t_0=1, t_p=0.5, t_2=2, t_{1a}=0.7,t_{1b}=1.2,r_{1a}=5,r_{1b}=7,\omega=5$, all eigen-modes exhibit non-zero values of IPR and NHSE can be observed given open boundary conditions (OBCs). (b) When the relative phase $\phi$ is altered to $\phi=0$,  NHSE will be switched off. (c) For the case that the relative phase $\phi=0$, the momentum space spectrum encloses zero area and $\mathcal{C}=0$. For the case $\phi=\frac{\pi}{2}$, NHSE can be present with  $\mathcal{C}\neq 0$. (d) The localization profile $\rho=|\psi|^2$ differs for Floquet eigen-state obtained with $\phi=0, |\epsilon|=3.36$ (blue lines), $\phi=\frac{\pi}{3}, |\epsilon|=0.89$ (brown lines), $\phi=\frac{\pi}{2}, |\epsilon|=1.51$ (yellow lines). Subfigure demonstrates that the maximum value of IPR changes with the variation of relative phase $\phi$.    } \label{fig1}
\end{figure*}
\begin{figure}[t]
	\centering
	\includegraphics[width=0.48\textwidth,height=0.57\textheight]{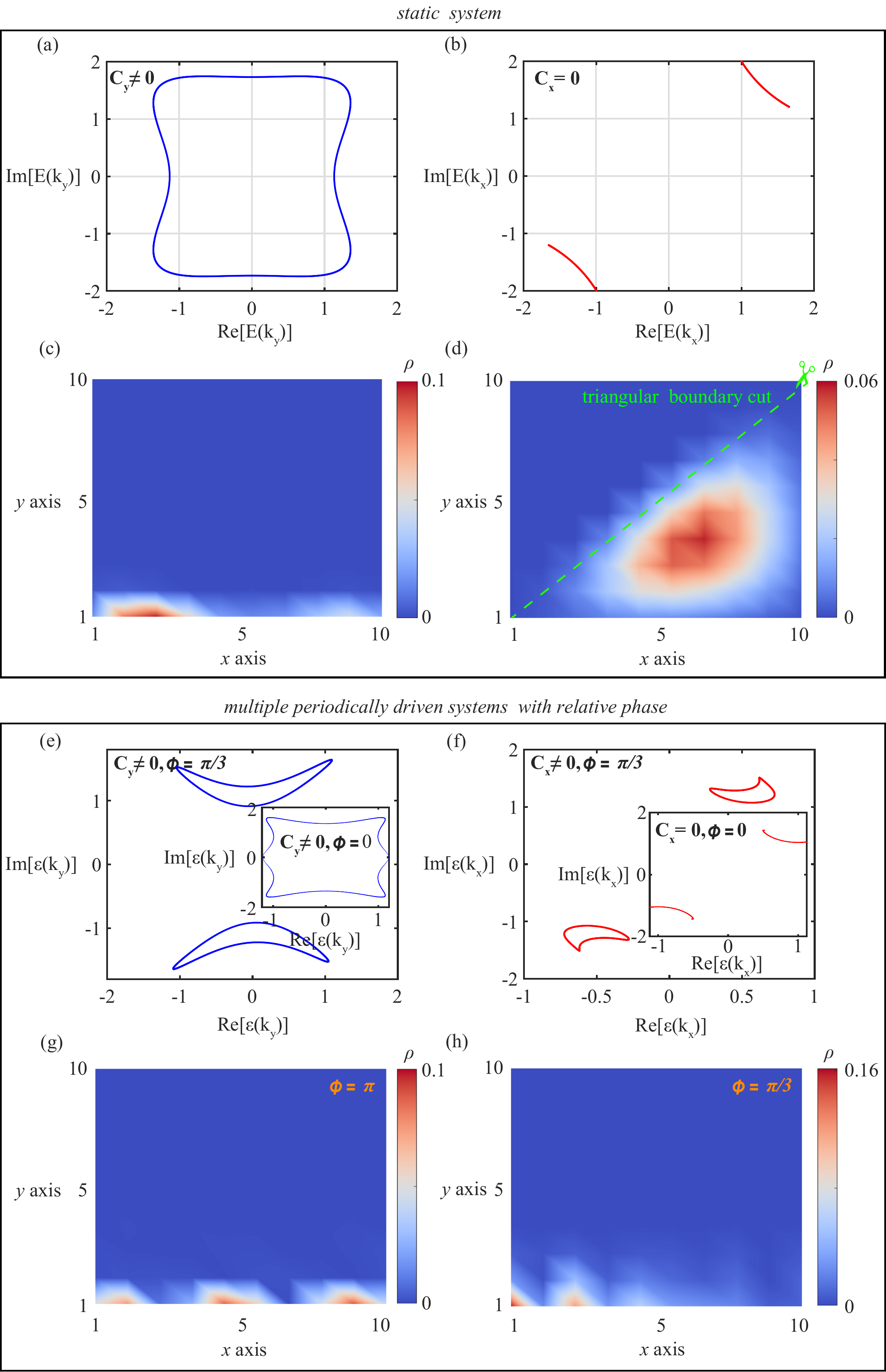}
	\caption{(a)-(b) For the static modified non-Hermitian QWZ model with $\gamma=2.0,v=1.0,\mathcal{M}=1.0$, non-zero winding formed by the momentum space energy spectrum $\{\mathrm{Re}[E(k_y)], \mathrm{Im}[E(k_y)]\}$ can be observed with fixed $k_x=1.5$ in (a), and for all values of fixed $k_y$, the momentum space spectrum  encloses zero area in (b). (c-d) Correspondingly, NHSE in the square shape boundary conditions can only be observed with the down-edge localization in the 2D lattice. For the triangular shape boundary conditions, NHSE will be manifested at the hypotenuse side. (e)-(f) With the multiple time periodic driving and $\mathcal{M}=1,\gamma_0=2,\gamma_1=1,v_0=1,v_1=2,\omega=3$, the direction favored skin localization persists for $\phi=p\pi,p\in Z$, and for the general case $\phi=\frac{\pi}{3}$, NHSE exits in both $x$ and $y$ directions with $C_x\neq 0, C_y\neq 0$. (g-h) For $\phi=\frac{\pi}{3}$, the real space skin localization depicted by $\rho=|\Psi|^2$, can be found at  the down-left edge. For $\phi=p\pi,p\in Z$, the y-direction favoured NHSE is preserved. } \label{fig2}
\end{figure}

\textit{Switching NHSE via tunable PT symmetry induced by  relative phase.---} NHSE in a non-Hermitian quantum chain can be typically  indicated by the shape of the generalized Brillouin zone, which is supposed to be characterized by polynomials that $f(\beta, E)=det[E-H(\beta)]$, $\beta$ being the generalized Bloch wave vector. To illustrate how NHSE can be switched during the dynamical process, we start by the static 1D quantum chain with PT symmetry, of which the constraints will result in the relation  $f(\beta, E)=f(\beta^{-1}, E)$ \cite{See.supplementary.materials.for.details}, and consequently no NHSE can be observed since $|\beta|=|\beta^{-1}|=1$. However, such prohibited  NHSE can emergently be reactivated upon exercising time-periodic driving. As an concrete illustration, we consider the bipartite quantum chain and the equation of motion for each sublattice can be described by
\begin{eqnarray}
\begin{aligned}
\dot{c}_{i,A}=&\gamma_2 c_{i,A}-i\gamma_1 c_{i,B}-i(t_1+t_2)c_{i+1,B}+i(t_1-t_2)\\
&c_{i-1,B}+(t_p-it_0)(c_{i+1,A}+c_{i-1,A}),
\end{aligned}
\end{eqnarray}
 \begin{eqnarray}
\begin{aligned}
\dot{c}_{i,B}=&-\gamma_2 c_{i,A}\!+\!i\gamma_1 c_{i,B}\!+\!i(t_1\!-\!t_2)c_{i-1,A}
-i(t_1-t_2)\\
&c_{i+1,A}-(t_p+it_0)(c_{i+1,B}+c_{i-1,B}),
\end{aligned}
\end{eqnarray}
where $c_{i,A(B)}$ denotes the annihilation of an $A(B)$ particle at $i$th site and $\gamma_{1}, -\gamma_1$ suggest the leftward and rightward intra-cell tunneling.  $\pm(t_1+t_2),\pm(t_1-t_2),(t_0\pm i t_p)$ depict the non-reciprocal inter-cell tunneling. $L$ is the length of the chain. Considering  the the spatial Fourier transformation, the momentum space Hamiltonian can be described as
\begin{eqnarray}
\begin{aligned}
\hat{H}_1(k)=&\sum_k (2t_0\cos{k})I+ (2it_p\cos{k}+i\gamma_2)\sigma_z \\
&+(2it_1\sin{k})\sigma_x+ (2it_2\cos{k}+i\gamma_1)\sigma_y.
\end{aligned}
\end{eqnarray}
According to Eq.(3), it can be seen that the PT symmetry is preserved, and there exists no global inversion symmetry. Correspondingly, it is supposed to be fulfilled that
\begin{equation}
(Q\kappa) \hat{H}_1 (k) (Q\kappa)^{\dagger}=H_1(k),
\end{equation}
where $Q=\sigma_y$ and $\kappa$ is the complex conjugate operator. Such a symmetry constraint is capable of prohibiting NHSE in the static quantum chain despite the presence of non-reciprocal tunneling  and  all  bulk eigen-states are uniformly distributed along the whole lattice. Meanwhile, the localization strength indicator, $\mathrm{IPR}=\frac{\sum_m |\psi_m|^4}{(\sum_m |\psi_m|^2)^2}$, will take vanishing value for all non-topological edge modes \cite{See.supplementary.materials.for.details}.

When the time-periodic driving is applied on the quantum tunneling $t_1=f(t), f(t+T)=f(t), \gamma_1=z(t), z(t)=z(t+T)$, there can be no well defined ground states as the system periodically exchanges particles and energies with the external driving field. Therefore, the proper topological characterization should be achieved based on the Floquet evolving operator defined as $U(k,T)= \mathcal{T} e^{-i\int^{T}_0 \hat{H}_1(k,t) dt}= e^{-i \hat{H}^s_{eff}(k) T},U(k,T)\Psi(k, t)= e^{-i \epsilon T}\Psi(k,t)$, and $\Psi(k,t)= e^{-i\epsilon t}\varphi(k, t), \varphi(k,t)=\varphi(k,t+T).$ Here, $\mathcal{T}$ depicts the time-ordering operator and $\Psi(t)$ is the Floquet eigen-state.  $H^s_{eff}$ represents the time-independent effective Hamiltonian.  It should be noticed that the PT symmetry constraints at each instantaneous time $\hat{H}_1(k,t)$ can not guarantee the PT symmetry of $U(T)$ (or $H^s_{eff})$.

To prohibit NHSE of Floquet states by PT symmetry in the periodically driven system, a sufficient condition of the dynamical driving can be listed as following
\begin{eqnarray}
\begin{aligned}
Q \hat{H}^{*}_1(k, t+t_{v}) Q^{\dagger}=\hat{H}_1(k, -t+t_v),
\end{aligned}
\end{eqnarray}
where $t_v$ is an arbitrarily chosen time point. To prove this, it should be noticed  that a PT symmetric time independent effective Hamiltonian is equivalent to set
$Q U^{*}(k,T) Q^{\dagger}= U^{-1}(k,T)$.  Meanwhile, based on Eq.(5), it can be derived that
\begin{eqnarray}
\begin{aligned}
&Q U^*(k,T) Q^{\dagger}= Q e^{i\Delta t H^{*}_1(k,T-\Delta t)}...e^{i\Delta t H^{*}_1(k,0)} Q^{\dagger}\\
&=e^{i \Delta t \hat{H}_1(k,2t_0\!-\!T+\Delta t)}e^{i\Delta t \hat{H}_1(k,2t_0\!-\!T\!+\!2\Delta t)}...e^{i\Delta t \hat{H}_1(k,2t_0)}\\
&=[e^{-i\!\Delta\!t \!\hat{H}_1(k,2t_0\!+\!T)}\! e^{-i\Delta\! t\! \hat{H}_1(k,2t_0\!+\!T-\!\Delta t)}...\!e^{-i\!\Delta t\!\hat{H}_1(k,2t_0\!+\!\Delta t)}]^{-1}\\
&=U(2t_0+T, 2t_0)^{-1}=U(T,0)^{-1}.
\end{aligned}
\end{eqnarray}
Specific to our model, when the dynamical driving is kept to the form $ t_1(t)=t_{1b}\cos{(\omega t)}+t_{1a}, r_1(t)=r_{1b}\cos{(\omega t) }+r_{1a}$, Eqs. (5) and (6) will be automatically satisfied by setting $t_v=0$. Numerical results in Figs. 1(a)-1(b) serve as a detailed illustration, where NHSE is totally forbidden.

To gain switchable NHSE, the relative phase $\phi$ between multiple time-periodic driving, which can not be gauged away through artificially choosing different initial time of driving, can be applied and the dynamical intra-cell tunneling  can be reassumed to the form $r_1(t)= r_{1b}\cos{(\omega t+\phi)}+r_{1a}$. Specifically, when the relative phase $\phi$ between $t_1(t)$ and $r_1(t)$ takes the value $\phi=2n\pi+\phi_0, \phi_0\neq p\pi, p\in Z$, the global PT symmetry of the Floquet operator is broken and the emergent NHSE can artificially be switched on. Meanwhile, since NHSE can be identified by the invariant, $\mathcal{C}=\int^{\pi}_{-\pi} \frac{dk}{2\pi i}\partial_k[\ln \det{(H^{s}_{eff}(k)-\epsilon_0)]}$, which is manifested by non-zero area enclosed by the momentum space Floquet energy spectrum  $\{\mathrm{Re}[\epsilon(k)], \mathrm{Im}[\epsilon(k)]\}$,  we demonstrate the phase transition regarding the presence of NHSE with different choice of $\phi$ [see Fig. 1(c)], and a detailed presentation of changes in localization profile is shown by Fig. 1(d). In this example, the relative phase $\phi$ manipulates NHSE by breaking the global temporal symmetry. Such a formalism can also be applied to system with $\mathcal{CP}$ symmetry in a similar way, but will fail for the cases of particle-hole symmetry (or time-reversal symmetry), where $U(T)$ or $H^{s}_{eff}$ will directly inherit the temporal symmetry constraints from the instantaneous time Hamiltonian \cite{See.supplementary.materials.for.details}.

\textit{Tunable localization direction of NHSE induced by relative phase in higher dimensions.---} Apart from being the switch of NHSE in the 1D quantum chain, the relative phase $\phi$ between multiple time periodic driving can also alter the localization direction of skin modes in 2D or higher dimensions. To illustrate this via a concrete example, we consider the modified dynamical non-Hermitian Qi-Wu-Zhang (NH-QWZ) model, where the Hamiltonian takes the form
\begin{eqnarray}
\begin{aligned}
H_{\mathrm{2D}}&=\sum_{k_x,k_y}[v(t)+\cos{k_x}+\cos{k_y}]\sigma_x+(\sin{k_x})\sigma_y\\
&+[\mathcal{M}\sin{k_y}+i\gamma(t)]\sigma_z.
\end{aligned}
\end{eqnarray}
The time-periodic driving is assumed to the form $\gamma(t)=\gamma_0+\gamma_1\cos{(\omega t+\phi)},v(t)=v_0+v_1\cos{(\omega t)}$. The instantaneous time Hamiltonian possesses (or does not possess) PT symmetry when $\mathcal{M}$ is purely imaginary (or real), and $[H_{\mathrm{2D}},\mathrm{PT}]=0, \mathrm{PT}=\sigma_x \kappa$. Consequently, for the static case and broken PT symmetry with $\mathrm{Im}(\mathcal{M})=0$, there can exist NHSE with favourable localization directions given different  boundary conditions. Specifically, skin modes can only be observed with Y-OBCs, and shall be absent for X-OBCs, Y-PBCs. In Figs. 2(a)-2(b), it is shown that  non-zero winding depicted by $\mathcal{C}_{x(y)}=\int^{\pi}_{-\pi} \frac{dk_{x(y)}}{2\pi i}\partial_{k_{x(y)}} \{\ln \det [H(k_{x(y)})-E_0]\}$ can be manifested with $\mathcal{C}_y\neq 0$ given $k_x=1.5$, and $\mathcal{C}_x=0$ for all values of fixed $k_y$. Correspondingly, in Fig. 2(c), the real space localization profile is demonstrated with the square-shape boundary conditions, and all the states are only localized at the down edge of the 2D lattice. In Fig. 2(d), the system is cut to suit the triangular-shape boundary, and the localization of skin modes is changed to the hypotenuse side. As the multiple time-periodic driving is turned on, such a direction-favoured NHSE can only persist for $\phi=2n\pi+p\pi$, and  considering changing the relative phase to $\phi=\frac{\pi}{3}$, skin modes are capable to  be  present in both $x$ and $y$ directions with X-OBCs and Y-OBCs, [see Figs. 2(e)-2(h)].


Here, how the relative phase $\phi$ manipulates  the localization direction of 2D-NHSE  with $\mathrm{Im}(\mathcal{M})=0$, can be attributed to a much different formalism in comparison to the 1D case. Specifically, the relative phase $\phi$ between multiple time periodic driving can dramatically alter the spatial structures of the 2D time independent effective Hamiltonian $H^s_{eff}$ and thus lead to diverse density pattern of skin modes. As a detailed illustration, we consider a temporal transition on the basis and $\tilde{\varphi}(t)=S(t)\varphi(t)$. If $S(t)=S(t+T)$ is satisfied, then it is fulfilled that $\tilde{\varphi}(t)=\tilde{\varphi}(t+T)$. According to the Floquet Schrodinger equation $(H_{2D}-i\partial_t)\varphi(t)=\epsilon(k,t)\varphi(t)$, the transformed state is supposed be governed by the following equations
\begin{eqnarray}
\begin{aligned}
&[\tilde{H}_{2D}(k,t)-i\partial_t]\tilde{\varphi}(t)=\epsilon\tilde{\varphi}(t),\\
& \tilde{H}_{2D}=S^{-1}(t)H_{2D}S(t)-i S^{-1}(t)\dot{S}(t).
\end{aligned}
\end{eqnarray}
It can be noticed that $\varphi(t)$ and $\tilde{\varphi}(t)$ share the same Floquet quasi-energy $\epsilon(k)$, and thus the transformed model and the original system exhibit the same features regarding the presence of NHSE. According to  the dynamical form of $v(t)$, we choose $S(t)=e^{-i\frac{v_1}{\omega}\sin{(\omega t)}\sigma_x},SS^{\dagger}=I$, and it can be obtained
\begin{eqnarray}
\begin{split}
&\tilde{H}_{2D}=[v_0+\cos{(k_x)}+\cos{(k_y)}]\sigma_x+\{\sin{(k_x)}\cos{[2pg(t)]}\\
&+[\sin{(k_y)}+i\gamma(t)]\sin{[2pg(t)]}\}\sigma_y+\{[\sin{k_y}+i\gamma(t)]\\
&\cos{[2pg(t)]}-\sin{(k_x)}\sin{[2pg(t)]}\}\sigma_z.
\end{split}
\end{eqnarray}
Here, $p=\frac{v_1}{\omega}$ and $g(t)=\sin{(\omega t)}$. To analytically  show how the spatial symmetry gets changed upon the time-periodic driving, we apply the  the Magnus expansion method and the time-independent effective Hamiltonian can be recast as
\begin{eqnarray}
\begin{aligned}
\tilde{\mathcal{H}}^{eff}_{2D} (k)&=\tilde{\mathcal{H}}^0_{2D}(k)+\frac{1}{\omega}[\tilde{\mathcal{H}}^{0}_{2D}(k),\tilde{\mathcal{H}}^{1}_{2D}(k)]-\frac{1}{\omega}[\tilde{\mathcal{H}}^{1}_{2D}(k),\\
&\tilde{\mathcal{H}}^{-1}_{2D}(k)]-\frac{1}{\omega}[\tilde{\mathcal{H}}^{-1}_{2D}(k), \tilde{\mathcal{H}}^{1}_{2D}(k)]+...,
\end{aligned}
\end{eqnarray}
where $\tilde{\mathcal{H}}^q_{2D}(k)=\frac{1}{T}\int^{T}_0 \tilde{\mathcal{H}}_{2D}(k,t)e^{-i q\omega t} dt$. In the high frequency limit $\omega\gg |\tilde{\mathcal{H}}^q_{2D}(k)|$, the approximation $\tilde{\mathcal{H}}^{eff}_{2D}(k)\approx \tilde{\mathcal{H}}^0_{2D}(k)$ will not change the point gap topological properties. Thus, given $\phi=2n\pi+p\pi, p\in Z$, the dynamical term like  $\gamma(t)\sin[2p g(t)]$ (or $\gamma(t)\cos[2p g(t)]$) will be odd (or even) function of time, which is supposed to be averaged to zero (or non-zero) value in $\tilde{\mathcal{H}}^{eff}_{2D}$. Consequently, it can be obtained that  $\tilde{\mathcal{H}}^{eff}_{2D}(k)=d^{eff}_{x}(k)\sigma_x+d^{eff}_y(k)\sigma_y+d^{eff}_z(k)\sigma_z$,   $d^{eff}_x(k)=[v_0+\cos{(k_x)}+\cos{(k_y)}],d^{eff}_y(k)=\sin{(k_x)}J_0(\frac{2v_1}{\omega}), d^{eff}_z(k)=\sin{(k_y)}J_0(\frac{2v_1}{\omega})+i\mathcal{N}$, and $\mathcal{N}=\frac{1}{T}\int^T_0\gamma(t)\cos{[2pg(t)]}dt$, where $J_0(x)$ is the zeroth order of Bessel function.  The time-independent effective Hamiltonian possesses totally different spatial structures (and the corresponding energy spectrum) in contrast to the instantaneous time Hamiltonian, where for fixed value of $k_{x_0}$, the Floquet eigen-energy will be decided by  $\epsilon^y(k)=\sqrt{[d^{eff}_x(k_y)]^2+[d^{eff}_z(k_y)]^2}$. Meanwhile, $d^{eff}_y(k_{x_0})$ can be deemed as a constant value, and both the real and imaginary part of $\epsilon^y(k)$ is $k_y$ dependent, which will consequently result in non-zero windings in the momentum space energy spectrum depicted by $C_y\neq 0$.  For the other case that $k_y$ is fixed, $\epsilon^x(k)=\sqrt{[d^{eff}_x(k_x)]^2+[d^{eff}_y(k_x)]^2}$ and $d^{eff}_z(k_{y_0})$ is a constant. It shall be noticed that in correspondence to the NHSE in $x$ direction, although $\epsilon=\sqrt{\epsilon^x (k)^2+[d^{eff}_z(k_{y_0})]^2}$ is complex, $\epsilon^x(k)$ shall be totally real and the zero winding number indicated by $\mathcal{C}_x=0$ is supposed to be fulfilled.    A change of relative phase from $\phi=2n\pi+p\pi$ to other general values (e.g. $\phi=\frac{\pi}{3}$) can greatly change the spatial structures, and $\epsilon^{x\, \mathrm{and}\, y}(k)$ contain both the real and the imaginary $k$-dependent part, where $\mathcal{C}_{x,y}\neq 0$ suggests the NHSE with no favourable directions.

\begin{figure}[t]
	\centering
	\includegraphics[width=0.5\textwidth,height=0.297\textheight]{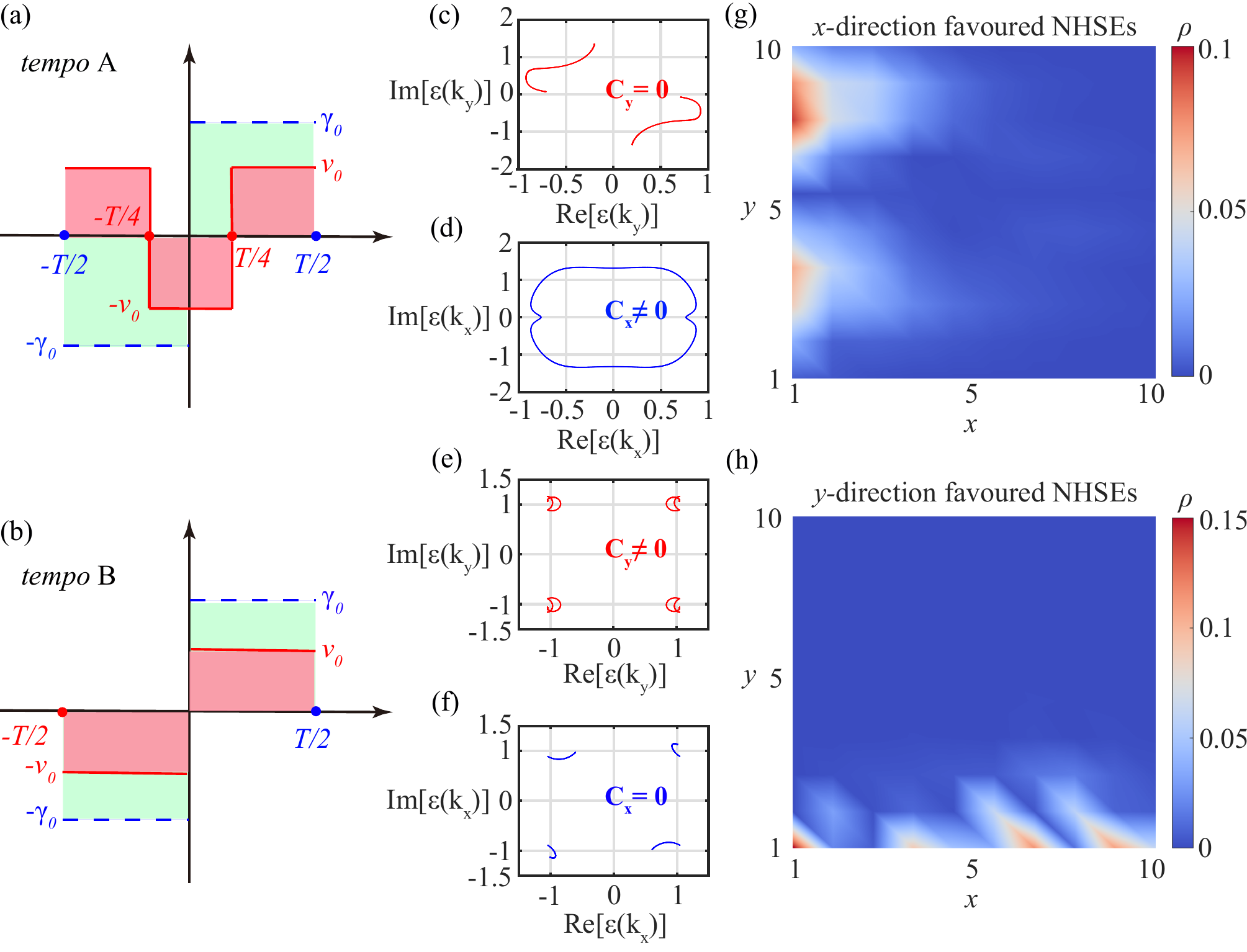}
	\caption{(a-b) The schematic picture of the quench driving formalism is shown, where the different (or same) tempo between the two quenches depicted by (a) (or (b)) is similar to the case of setting relative phase $\phi=\frac{\pi}{2}$ (or $\phi=0$). (c-d) In correspondence to the quench dynamics in (a) and  $T=3,\gamma_0=2,v_0=1$, NHSEs can only be observed in $x$ direction with $C_y\neq 0$ and fixed $k_x=0.3$. For all values of fixed $k_y$, it is demonstrated $C_x=0$. (e-f) When the quench formalism is changed to (b), localization direction of NHSEs gets altered and $C_y\neq 0, C_x=0$.  (g) For the same parameters in (a) and X-OBCs, Y-OBCs, the Floquet state with $\epsilon=-0.5169-1.3071i$ is localized at the left edge of the 2D lattice. (h) Upon changing to the driving formalism in (b) and X-OBCs, Y-OBCs, the Floquet state turns to be localized at the down edge. } \label{fig3}
\end{figure}

How the relative phase between multiple time-periodic driving affects the localization direction of NHSE is also greatly related to the exact details of the dynamical driving, and in contrast to the cosinusoidal type driving formalism, results may become more fruitful upon considering the quench process. To reveal this, it shall be noticed that the different relative phase can be similarly achieved by setting different tempos between the multiple quench dynamics and it is reassumed $v=v_0, \gamma=\gamma_0$ for $t\in [\frac{T}{4},\frac{T}{2}]$, $v=-v_0,\gamma=\gamma_0$ for $t\in [0,\frac{T}{4}]$, $v=-v_0,\gamma=-\gamma_0$ for $t\in [-\frac{T}{4},0]$, and $v=v_0,\gamma=-\gamma_0$ for $t\in [-\frac{T}{2},-\frac{T}{4}]$.  This dynamical process is similar to set $\phi=\frac{\pi}{2}$, which can be termed as tempo A. Meanwhile,  we may also retake $v=v_0,\gamma=\gamma_0$ for $t\in [0,\frac{T}{2}]$, $v=-v_0,\gamma=-\gamma_0$ for $t\in[-\frac{T}{2},0]$, in correspondence to relative phase $\phi=0$ (termed as tempo B), where the two quenches are of the equal pace.  A schematic picture of the driving  formalism is shown in Figs.3(a)-3(b).  Numerically, it is demonstrated that  the direction-favoured NHSE can be maintained for both two tempos given $\mathrm{Im}(\mathcal{M})=0$, and remarkably the localization direction of skin modes  will be interchanged upon altering from tempo A to tempo B [see Figs.3(c)-3(h)].


Based on the discussion above, it can be seen that the relative phase can turn  NHSE of the Floquet states by breaking the temporal symmetry constraints or changing the spatial structures of time-independent effective Hamiltonian, where the two formalisms can also be combined together to decide the skin localization patterns.  Specifically, considering the static NH-QWZ model with $\mathrm{Re}(E)=0$,  PT symmetry will forbid all kinds of NHSEs. As the multiple time-periodic driving is considered, the direction favored NHSE can not be observed with tempo B in the quench process (or $\phi=2n\pi+p\pi$ in the cosinusoidal type driving) due to the preserved  PT symmetry of $U(T)$. For the other general case, skin localization will be altered with the variation of spatial structures of $H_{eff}$ obtained with different driving tempos. For example, we term the dynamical quench process, where $v=v_0,\gamma=-\gamma_0$ for $t\in [-\frac{T}{2},-\frac{T}{12}]$, $v=-v_0,\gamma=-\gamma_0$ for $t\in [-\frac{T}{12},0]$, $v=-v_0,\gamma=\gamma_0$ for $t\in [0,\frac{5T}{12}]$, and $v=v_0,\gamma=\gamma_0$ for $t\in [\frac{5T}{12},\frac{T}{2}]$, as tempo C.  The skin localization can be altered from $C_x\neq 0, C_y=0$ to $C_y\neq 0, C_x\neq 0$ upon changing from tempo A to tempo C.

\textit{Discussion.---} We have demonstrated how to control the emergence and density pattern of NHSE by manipulating the relative phase between the multiple time periodic driving. Such a mechanism can be feasibly realized and observed by shaking dissipative optical lattice in the real experiments \cite{NJP,PhysRevLett.109.145301,PhysRevA.102.023328,PhysRevA.89.061603}, where for the 1D case, the two sublattice can be implemented by the dynamical supper lattice potential of the form $V(x)=V_{xs} \cos^2{(\pi x/a_x)}+V_{xl}\cos^2{(\pi x/2a_x+\chi (t))}$, and the driving frequency of $\chi(t)$ shall be set to resonant with the characteristic frequency of the trapped atom so as to induce non-Hermiticity. Besides, the non-reciprocal inter-cell tunneling can be artificially achieved by combining the laser-assisted tunneling technique and the post-selection formalism given the conditions of large detuning \cite{PhysRevLett.111.185301,PhysRevLett.111.185302,Adv.Phys.63.77}. The tunneling amplitude is proportional to the two-photon Rabi frequency, of which the dynamical features together with the relative phase can be tuned with the application of acousto-optic modulator.  Meanwhile, our results on dynamical NHSE can also be applied and simulated in other diverse platforms exhibiting Floquet dynamics, such as the discrete time quantum walk \cite{Natphysd,PhysRevA.81.042330} and circuit lattices \cite{PhysRevLett.119.093901,journalofappliedphysics,6n5m-wvs7,PhysRevA.105.042211,PhysRevA.99.012333}. In conclusion, our works shed light on finding new ways to realize the tunable NHSE in various dimensions and reveal how the relative phase decides the favorable direction of NHSE.

\textit{Acknowledgments---} Huan-Yu Wang acknowledges support from the start up funding of Fuzhou University under Grant XRC-23079, and  the Fujian Province Young and Middle-aged Teachers Education Research Project (Science and Technology Category) under Grant No. JAT251004.

\end{document}